\documentclass[%
 aip,
 jmp,%
 amsmath,amssymb,
 reprint,%
]{revtex4-1}

\usepackage{graphicx}
\usepackage{dcolumn}
\usepackage{bm}

\begin{document}

\title{Transition Edge Sensors with Few-Mode Ballistic Thermal Isolation}

\author{D. Osman}
 \email{d.osman@mrao.cam.ac.uk}
\author{S. Withington}
\author{D. J. Goldie}
\author{D. M. Glowacka}
\affiliation{Quantum Sensors Group, Cavendish Laboratory, Cambridge, United Kingdom CB3 0HE}


\begin{abstract}
We have fabricated Transition Edge Sensors (TESs) whose thermal characteristics are completely characterised by few-mode ballistic phonon exchange with the heat bath. These TESs have extremely small amorphous SiN$_{\rm x}$ support legs: 0.2~$\mu$m thick, 0.7 to 1.0~$\mu$m wide and 1.0 to 4.0~$\mu$m long. We show, using classical elastic wave theory, that it is only necessary to know the geometry and bulk elastic constants of the material to calculate the thermal conductance and fluctuation noise. Our devices operate in the few-mode regime, between 5 and 7 modes per leg, and have noise equivalent powers (NEPs) of 1.2~aW~Hz$^{-1/2}$. The NEP is dominated by the thermal fluctuation noise in the legs, which itself is dominated by phonon shot-noise. Thus TESs have been demonstrated whose thermal characteristics are fully accounted for by an elastic noise-wave model. Our current devices, and second-generation devices based on patterned phononic filters, can be used to produce optically compact,  mechanically robust, highly sensitive TES imaging arrays, circumventing many of the problems inherent in conventional long-legged designs.
\end{abstract}


\maketitle

\section{Introduction}

The sensitivity of a well-designed Transition Edge Sensor (TES) is determined by the thermal fluctuation noise in the weak link to the heat bath \cite{Irwin2005}. Noise-equivalent powers (NEPs) as low as 0.5~aW~Hz$^{-1/2}$ are now routinely achieved \cite{Goldie2011, Beyer2011, Benford2004, Khosropanah2000,Audley2012,Khosropanah2012a}. Our own ultra-low-noise TESs, developed for the SAFARI instrument on SPICA \cite{Goldie2012,Goldie2012a}, have four SiN$_{\rm x}$ support legs, 200~nm thick, 2.1~$\mu$m wide and 540~$\mu$m long, giving a total thermal conductance $G$ of 0.19~pW~K$^{-1}$ and an NEP of 0.42~aW~Hz$^{-1/2}$. The usual approach to describing heat flow in the legs is to use $P(T_{h},T_{c}) = K (T^{n}_{h} - T^{n}_{c})$, where $T_{h}$ and $T_{c}$ are the hot and cold termination temperatures respectively. Once the values of $K$ and $n$ have been determined through experiment, the conductance can be calculated, $G = n K T^{n-1}_{h}$, which in turn leads to the sensitivity through NEP$ = ( \gamma 4 k T_{h}^{2} G )^{1/2}$. $\gamma$ is a factor that accounts for the termination temperatures being different. This parametric approach is effective when designing TESs, but is inefficient to use because of the need to find $K$ and $n$ empirically, and gives little insight into the scattering mechanisms that determine heat flow and noise.

During the course of our TES development work, we have encountered a number of problems that are not easily solved when using conventional designs: (i) The conductance falls as $L^{-1}$, where $L$ is the length, and so ultra-low-noise devices ($<$ 0.2~aW~Hz$^{-1/2}$) require extreme geometries (200~nm thick, 1~$\mu$m wide and 1.0~mm long), which lead to manufacturing and operational problems. In the case of far-infrared imaging arrays, it is difficult to achieve close optical packing even when large-aperture lightpipes are used. (ii) The variation in thermal conductance across notionally identical devices in an array is typically $\pm$~15\%, which is the limiting factor in uniformity of performance. It is likely that this wide variation, which cannot be accounted for in terms of simple geometrical or surface-roughness arguments, is due to phonon localisation in the disorder of amorphous SiN$_{\rm x}$. The scattering process that gives rise to the needed $L^{-1}$ dependence is likely to be the same as the process that leads to localisation, and so the variation in conductance is intrinsic to the material.\cite{Withington2011,Withington2013} (iii) Two level systems (TLSs) in SiN$_{\rm x}$ result in a specific heat that is 100 times greater than the Debye value, and therefore large volumes of dielectric, including long-narrow legs, lead to response times that are too slow for some applications. In the case of ultra-low-noise devices, the readout corner frequency can be as low as 10~Hz, making measurement and effective usage difficult because of $1/f$ noise. (iv) As the thermal fluctuation noise is reduced to very small values by increasing $L$, the measured NEP seems to deviate from the calculated value by a factor of about 1.2. The reason is uncertain, but the formula NEP $ = ( \gamma 4 k T_{h}^{2} G )^{1/2}$ does not take into account the nature of the scattering that determines $G$. Simulations show that inelastic scattering behaves in a different way to elastic scattering\cite{Withington2012}, and that inelastic scattering gives rise to seemingly increased noise levels due to the random exchange of energy between the internal degrees of freedom that constitute the losses and the travelling elastic waves.

In an attempt to overcome problems of this kind, we are investigating the use of micromachined phononic support legs to establish the weak thermal link to the heat bath. In this paper, we show that low-noise operation can be achieved by using few-mode ballistic support legs, and that it is not necessary to rely on a $1/L$ dependence to give low thermal conductances. The overall vision is to create a TES technology that allows compact optical designs; that avoids phonon localisation caused by an uncertain scattering mechanism, leading to more uniform arrays; that minimises the heat capacity, leading to faster detectors; and that uses macroscopically engineered elastic scattering to ensure that 
the NEP that falls as $G^{-1/2}$ as the elastic-wave transmission factor is reduced.

A key question is whether it is possible to manufacture TESs having legs that can only support a few elastic modes, and then if the legs are made short enough to ensure ballistic transport, whether the electro-thermal behaviour is consistent with calculations based on elastic noise-wave theory. If this cannot be done, it is unlikely that it will be possible to produce more complicated patterned components having predictable characteristics. In this paper we describe experimental work on a set of MoAu TESs having exceedingly short, narrow legs: 200~nm thick, 0.7 to 1.0~$\mu$m wide and 1.0 to 4.0~$\mu$m long. We show that it is possible to calculate the conductance and thermal fluctuation noise accurately from first principles, using only bulk elastic constants, and without the need for parametric models. We show that the NEP is already good enough for most submillimetre-wave and far-infrared astronomical instruments, even before phononic structures are introduced. On the basis of this work we discuss the potential benefits of using phononic filters in the legs of ultra-low-noise TESs.

\section{\label{sec:1}Theory}

To model the behaviour of few-mode ballistic TESs, we first calculated the dispersion relationships of the elastic waves that can propagate along the support legs, and then used Bose-Einstein statistics to determine the net heat flux and thermal fluctuation noise.

\subsection{\label{subsec:1}Elastic Waves}

The elastic wave equation is\cite{Timoshenko1986}
\begin{equation}
\rho\omega^{2}u_{i}+C_{ijkl}\frac{\partial^{2}u_{k}}{\partial x_{j}\partial x_{l}}=0
\mathrm{,}
\label{eq:1}
\end{equation}
where $u_{i}$ is the displacement field in Cartesian direction $i$, $C_{ijkl}$ is the fourth-rank stiffness tensor, $\rho$ is the mass density, and $\omega$ the angular frequency. Equation (\ref{eq:1}) can be solved by expressing the $u_{i}$ as weighted linear combinations of the basis-functions $\psi_{r}$,
\begin{equation}
u_{i}= a_{ir} \psi_{r}
\mathrm{.}
\label{eq:2}
\end{equation}
$a_{ir}$ is the $r$'th expansion coefficient of the $i$-directed displacement, and we have used the same basis functions for each direction. Thus for each vector component of the displacement field, there exists a complete, but not necessarily orthogonal, set of basis functions that spans the volume $V$ and characterises behaviour. Substituting Eq.~(\ref{eq:2}) into Eq.~(\ref{eq:1}), multiplying by the conjugate of $\psi_{r'}$ and integrating over $V$ leads to
\begin{equation}
\rho \omega^{2} a_{ir} \intop_{V} \psi_{r'}^{*} \psi_{r} {\rm d}V - a_{kr} C_{ijkl} \intop_{V} \psi_{r'}^{*} \frac{\partial^{2} \psi_{r}} {\partial x_{j} \partial x_{l}} {\rm} {\rm d}V = 0
\mathrm{.}
\label{eq:3}
\end{equation}

Integrating the second term on the left by parts, and noting that the traction on every free surface is zero, gives
\begin{equation}
\rho \omega^{2} \delta_{ik} a_{kr} \intop_{V} \psi_{r'}^{*} \psi_{r} {\rm d}V - a_{kr} C_{ijkl}
\intop_{V} \frac{\partial \psi_{r'}^{*}}{\partial x_{j}} \frac{\partial\psi_{r}}{\partial x_{l}}
{\rm d}V = 0
\mathrm{.}
\label{eq:4}
\end{equation}
Now define
\begin{equation}
R_{ir',kr} = \rho\delta_{ik} \intop_{V} \psi_{r'}^{*} \psi_{r} {\rm d}V
\mathrm{,}
\label{eq:5}
\end{equation}
and
\begin{equation}
S_{ir',kr} = C_{ijkl} \Gamma_{r'r,jl} = C_{ijkl} \intop_{V} \frac{\partial \psi_{r'}^{*}}{\partial x_{j}}
\frac{\partial\psi_{r}}{\partial x_{l}} {\rm d}V
\mathrm{,}
\label{eq:6}
\end{equation}
such that Eq.~(\ref{eq:4}) can be written in matrix form
\begin{equation}
\left[\omega^{2}\mathbf{R}-\mathbf{S}\right]  \mathbf{a} = {\bf 0}
\mathrm{,}
\label{eq:7}
\end{equation}
which is a generalized eigenvalue equation. Equation (\ref{eq:7}) can be solved numerically to  give the mode coefficients $\mathbf{a}$ corresponding to the frequency $\omega$. Once the mode coefficients are known, the spatial forms of the modes can be calculated through Eq.~(\ref{eq:2}).

Equation~(\ref{eq:7}) holds for any lossless material, but for an isotropic medium, such as an amorphous dielectric, the stiffness tensor is given by
\begin{equation}
C_{ijkl}=\lambda\delta_{ij}\delta_{kl}+\mu(\delta_{ik}\delta_{jl}+\delta_{jk}\delta_{il})
\mathrm{,}
\label{eq:8}
\end{equation}
where $\delta_{ij}$ is the Kronecker delta and the Lam\'{e} parameters,  $\lambda$ and $\mu$,  define the elastic properties. In practice, only the Young's modulus $E$ and Poisson's ratio $\nu_{p}$ are needed to calculate $\lambda$ and $\mu$. Substituting Eq.~(\ref{eq:8}) into Eq.~(\ref{eq:6}) leads to the convenient expression
\begin{equation}
S_{ir',kr} = \lambda \Gamma_{r'r,ik} - \mu \delta_{ik} \Gamma_{r'r,jj} - \mu \Gamma_{r'r,ki}
\label{eq:9}
\end{equation}
which aides the calculation of Eq.~(\ref{eq:7}).

\begin{figure}[h]
\noindent \begin{centering}
\includegraphics[scale=1]{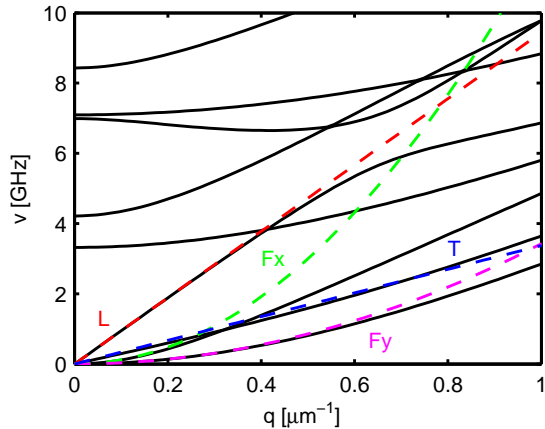}
\par\end{centering}
\caption{Dispersion relationships, frequency $\nu$ against wavenumber $q$, of a rectangular SiN$_{\rm x}$ bar having a thickness of 200~nm and a width of 700~nm. The dashed lines show analytic approximations to the four lowest order modes: longitudinal (red,L), torsional (blue,T), in-plane flexural (green,Fx) and out of plane flexural (mauve,Fy).
\label{fig1}}
\end{figure}

Various forms can be used for the expansion functions. We have used both Gaussian-Hermite functions and power series for structures having rectangular cross sections, but here we follow the approach of Nishiguchi\cite{Nishiguchi1997} and write
\begin{equation}
\psi_{r}(x,y,z) = \left(\frac{2x}{W}\right)^{m} \left(\frac{2y}{H}\right)^{n} \exp(iqz)
\mathrm{,}
\label{eq:10}
\end{equation}
where the index of each basis function $\psi_{r}$ represents a unique combination of the integers $m$ and $n$. $W$ and $H$ are the width and thickness of the leg along the $x$ and $y$ axes respectively. Travelling waves are assumed along the length of the leg, in the $z$ direction, and $q$ is the wave number.

To calculate a set of modal dispersion curves, it is only necessary to know $E$, $\nu_{p}$, $\rho$, $W$ and $H$. At each step, a value of $q$ is assumed, and then Eq.(\ref{eq:10}), Eq.(\ref{eq:5}), Eq.(\ref{eq:6}) and Eq.(\ref{eq:9}) are used to establish Eq.(\ref{eq:7}). The eigenvalues of Eq.~(\ref{eq:7}) give the modal frequencies, and the eigenvectors give the spatial forms of the displacements.

\begin{figure*}[t]
\noindent \begin{centering}
\includegraphics[scale=1]{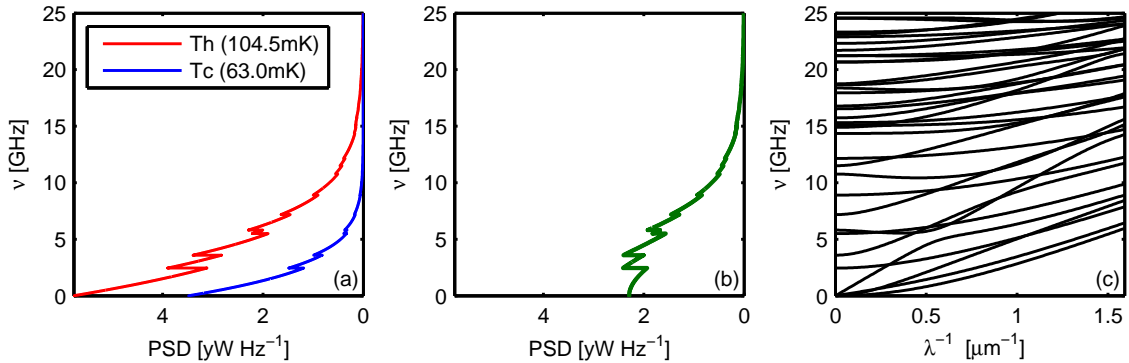}
\par\end{centering}
\caption{(a) Power spectral density (PSD) of the power flowing from the hot to the cold end (red), and cold to the hot end (blue), of a SiN$_{x}$ leg having a cross-section of 830~nm $\times$ 200~nm: Th=104.5mK and Tc=63mK. (b) PSD of the net power flowing along the leg. (c) Dispersion profiles of the elastic modes carrying the energy. The plots are oriented so that it can be seen how the individual modes contribute to the total flux.
\label{fig2}}
\end{figure*}

Figure~\ref{fig1} shows calculated dispersion profiles of a set of low-order elastic waves in an amorphous SiN$_{\rm x}$ leg having $H=$~200~nm and $W=$~700~nm. For amorphous SiN$_{\rm x}$\cite{Walmsley2007,Vlassak1992,Shackelford2001}, we used $E=$ 280 GPa, $\nu_{p}=$ 0.28 and $\rho=$ 3.14 g~cm$^{-3}$ . The precise forms of the modes were found to be relatively insensitive to the exact values used. In Fig.~\ref{fig1} we also plot analytic approximations to the lowest four dispersion relationships. These correspond to one longitudinal (dashed red), one torsional (dashed blue) and two flexural modes (dashed green and dashed mauve). Notice that they do not cut off, and so will propagate at all frequencies in a leg of any size. The analytic approximations used are as follows: (i) The longitudinal mode has a linear dispersion relationship given by $\omega = \sqrt{(E/\rho)} q$. (ii) The torsional mode is approximated by the expression, $\omega = 2 q \sqrt{(C_{44}/\rho)}(W/H)$, where $C_{44}$ is the $\{4,4\}$ element of the stiffness tensor in standard truncated notation. (iii) The two flexural modes are given by $\omega=q^{2}\sqrt{{EW^{2}}/{12\rho}}$ in the $x$ direction, and $\omega = q^{2} \sqrt{{EH^{2}}/{12\rho}}$ in the $y$ direction. The numerical and analytical calculations agree well in the small wavenumber limit.

\subsection{Heat Flux}\label{sec:p}

Once the dispersion relationships are known, it is possible to calculate the net thermal power travelling along the leg. In the case of a TES, one end of each leg is held at the bath temperature $T_{c}$, and the other end is held at the superconducting transition temperature $T_{h}$ by electrothermal feedback. Given that the TES island and Si carrier wafer are much larger than the bridge, we assume that the phonon occupancies of the counter-propagating waves are given by Bose-Einstein statistics: $f(\nu,T) = [\exp (h\nu/kT)-1]^{-1}$. The net power spectral density flowing from the TES island to the bath is then $PSD = \sum_{i} \left[ {h\nu}{f(\nu,T_{h})} - {h\nu}{f(\nu,T_{c})} \right]$. The sum is over the number of propagating modes at frequency $\nu$. The total heat flux is
\begin{equation}
P(T_{h},T_{c}) = \sum_{i} \left[ \int_{\nu_{i}}^{\infty} h \nu {f(\nu,T_{h})} d \nu - \int_{\nu_{i}}^{\infty} h \nu {f(\nu,T_{c})} {\rm d} \nu \right]
\mathrm{,}
\label{eq:11}
\end{equation}
where $\nu_{i}$ is the cut-off frequency of mode $i$. The total power transmitted by each mode is bounded on the low-frequency side by the cut-off frequency and on the high-frequency side by the roll off in occupancy. Figure \ref{fig2} shows elements of a complete calculation for a SiN$_{\rm x}$ leg having a cross section of  200~nm  $\times$ 830~nm: (a) The power spectral density of the flux flowing from the hot ($T_{h} =$ 104.5 mK) to the cold ($T_{c}=$ 63.0 mK) end (in red), and from the cold to the hot end (in blue). The discontinuities indicate step changes in power where modes cut on. (b) The net flux flowing along the microbridge. (c) The dispersion curves of the elastic modes on the structure. On inspection, it seems that for a 100 mK bath temperature, most of the power is carried by phonons having frequencies below 10 GHz.

\begin{figure}
\noindent \begin{centering}
\includegraphics[scale=1]{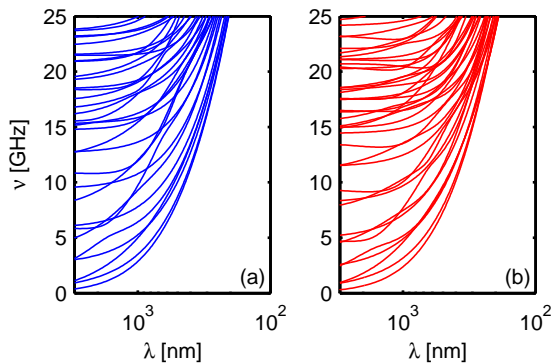}
\par\end{centering}
\caption{(a) Dispersion relationships for SiN$_{\rm x}$, where $\nu$ is the frequency and $\lambda$ is the wavelength. (b) Dispersion relationships for crystalline Si. Both plots are calculated for a rectangular leg having a cross section of 200~nm $\times$ 700~nm.
\label{fig3}}
\end{figure}

Figure \ref{fig3} shows the dispersion curves of (a) amorphous SiN$_{\rm x}$ and (b) crystalline Si legs having cross sections of 200~nm $\times$ 700~nm, but where we plot frequency as a function of wavelength rather than wavenumber. The stiffness of Si is anisotropic, and so it is necessary to calculate the dispersion profiles using Eq.~(\ref{eq:6}) rather than Eq.~(\ref{eq:9}): we used $C_{11}$=165.6~GPa, $C_{12}$=63.9~GPa and $C_{44}$=79.5~GPa \cite{Hopcroft2010}. Figure \ref{fig3} illustrates two points: (i) The two materials behave in a very similar way, and either could be used for making few-mode ballistic TESs. (ii) The majority of the power is carried by phonons having wavelengths greater than about 400~nm. We shall see later that our micro-fabricated devices are smooth on these scale sizes, and so we do not expect severe reflections at the positions where the ends of the leg meet the TES island and Si carrier wafer.

\subsection{Surface Roughness and Specular Reflection}

\begin{figure}[h]
\noindent \begin{centering}
\includegraphics[scale=1]{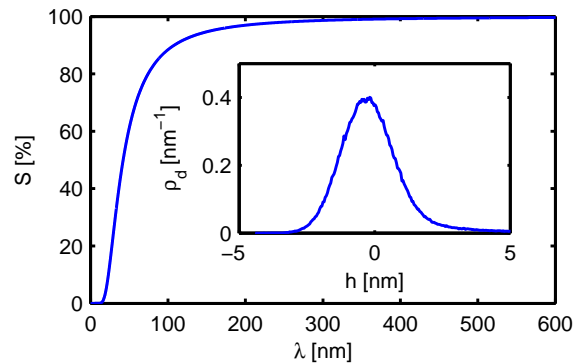}
\par\end{centering}
\caption{Specularity $S(\lambda)$ of SiN$_{\rm x}$ as a function of phonon wavelength, $\lambda$, calculated from a set of AFM measurements. The insert shows the probability density distribution of the surface roughness of all AFM measurements; an r.m.s. variation of $\pm 1.58$~nm is seen}
\label{fig4}
\end{figure}

In order for the thermal flux to attain the ballistic limit, the surface must be sufficiently smooth to guarantee specular reflection along the whole length of the structure. To assess the likelihood of diffusive scattering, we studied the surfaces of SiN$_{\rm x}$ membranes using Atomic Force Microscopy (AFM). We measured both the front and back surfaces of a set of membranes after they had been processed in exactly the same way as the TESs themselves. We took 20 $\mu$m $\times$ 20 $\mu$m scans, and calculated various statistical quantities, including the specularity:
\begin{equation}
S(\lambda) = \int_{0}^{\lambda / 2} p(\eta) \exp \left[ - \frac{16 \pi^{3} \eta^{2}}{\lambda^{2}} \right] {\rm d} \eta
\mathrm{,}
\label{eq:12}
\end{equation}
where $\lambda$ is the phonon wavelength, $\eta$ is the r.m.s. surface roughness of a given subdomain, and $p(\eta)$ is the probability distribution of $\eta$ over a set of subdomains. For each wavelength, a 20~$\mu$m $\times$ 20~$\mu$m image was divided into $2 \lambda \times 2 \lambda$ cells, and these formed the subdomains for the specularity calculations. The specularity essentially gives the probability that an incident phonon having wavelength $\lambda$ will be specularly reflected when a plane elastic wave is incident normally on the surface. The derivation is outlined in Ziman\cite{Ziman1960}, and it should be noted that because normal incidence is assumed, Eq.~(\ref{eq:12}) is a lower limit. Figure \ref{fig4} shows the specularity as a function of phonon wavelength calculated using the AFM measurements. For $\lambda>400$~nm, the probability that a phonon is specularly reflected approaches 100\%, and because in Sec$.$ \ref{sec:p} we saw that most of the energy is carried by phonons having $\lambda>400$~nm, it seems that elastic diffusive surface scattering can be ignored. The insert in Fig. \ref{fig4} shows the probability density distribution of the surface roughness of all measurements, and an r.m.s. variation of $\pm 1.58$~nm is seen.

\begin{figure*}[t]
\noindent \begin{centering}
\includegraphics[scale=1.0]{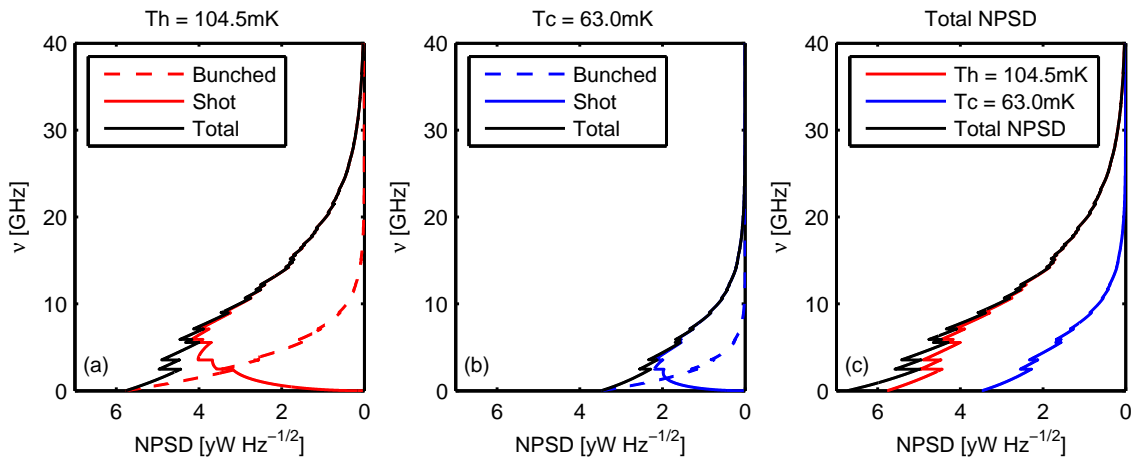}
\par\end{centering}
\caption{Noise power spectral density (NPSD) for a SiN$_{\rm x}$ leg having a cross section of 200~nm $\times$ 830~nm:
$T_{h}=104.5$ mK and $T_{c}=63.0$ mK. We show the bunched, shot and total fluctuation noise associated with (a) waves travelling from the hot to the cold end, (b) waves travelling from the cold to the hot end, (c) the net flux.}
\label{fig5}
\end{figure*}

\subsection{Thermal Fluctuation Noise}

Assuming ballistic transport, and Bose-Einstein statistics, it is possible to calculate the fluctuation in the absorbed power when the traveling elastic waves dissipate their energy in the terminations. Strictly, we are assuming that the leg is acoustically lossless, and that there is no scattering between the individual modes. The r.m.s. fluctuation in the absorbed power for waves traveling in one direction is given by
\begin{eqnarray}
\Delta P(T)^{2} = \sum_{i} 2B \int_{\nu_{i}}^{\infty} \left[ h\nu f(\nu,T) \right]^{2} d \nu \nonumber \\
+ \sum_{i} 2B \int_{\nu_{i}}^{\infty} \left[ h\nu \right]^{2} f(\nu,T) {\rm d} \nu
\mathrm{,}
\label{eq:13}
\end{eqnarray}
where $B$ is the readout bandwidth. The first term on the right hand side corresponds to classical bunched noise, and the second term to shot noise. Because the forward and backward traveling waves are uncorrelated, the total fluctuation is the quadrature sum of $\Delta P(T_{h})$ and $\Delta P(T_{c})$. 

In the case of a well-designed TES, the fundamental limit to the Noise Equivalent Power (NEP) is the fluctuation in the heat flow in the legs, and therefore we can directly interpret $\Delta P$, with $B=1$~Hz, as the NEP. Figure \ref{fig5} shows the noise power spectral density (NPSD) of the fluctuations for a SiN$_{\rm x}$ microbridge having a cross section of 200~nm $\times$ 830~nm. (a) shows the fluctuations in the forward direction, (b) in the reverse direction, and (c) the quadrature sum. The plots indicate that at these temperatures, the NEP is dominated by phonon shot noise.

\section{\label{sec:2}Experiment}

Our TESs comprised a 70~$\mu$m $\times$ 70~$\mu$m MoAu bilayer, having a critical temperature of around 105~mK, on a 200~nm thick amorphous SiN$_{\rm x}$ membrane. The electrical readout leads, along two of the legs, were made of Nb. We have developed this technology extensively for ultra-low-noise (NEP $\approx$ 0.4~aW~Hz$^{-1/2}$) detectors for astronomy, but we have not used it previously with ballistic support legs. The TESs were read out using two-stage SQUIDs \cite{Drung2007} having a noise current of approximately 8 pA~Hz$^{-1/2}$. Full electrothermal TES simulations were performed to design the devices. The TESs and SQUIDs were mounted in a blackened, light-tight box, and placed inside an Adiabatic Demagnetisation Refrigerator (ADR). Regulating the current in the ADR magnet allowed the temperature of the sample stage to be controlled, down to a minimum of 60~mK. By taking IV measurements of the TESs, we were able to determine the power flowing through the legs as a function of bath temperature. We were particularly careful to calibrate the system, for example measuring the value of the TES bias resistor. These well-known techniques have been described previously \cite{Goldie2011}.

\begin{figure}[h]
\begin{centering}
\includegraphics[scale=1]{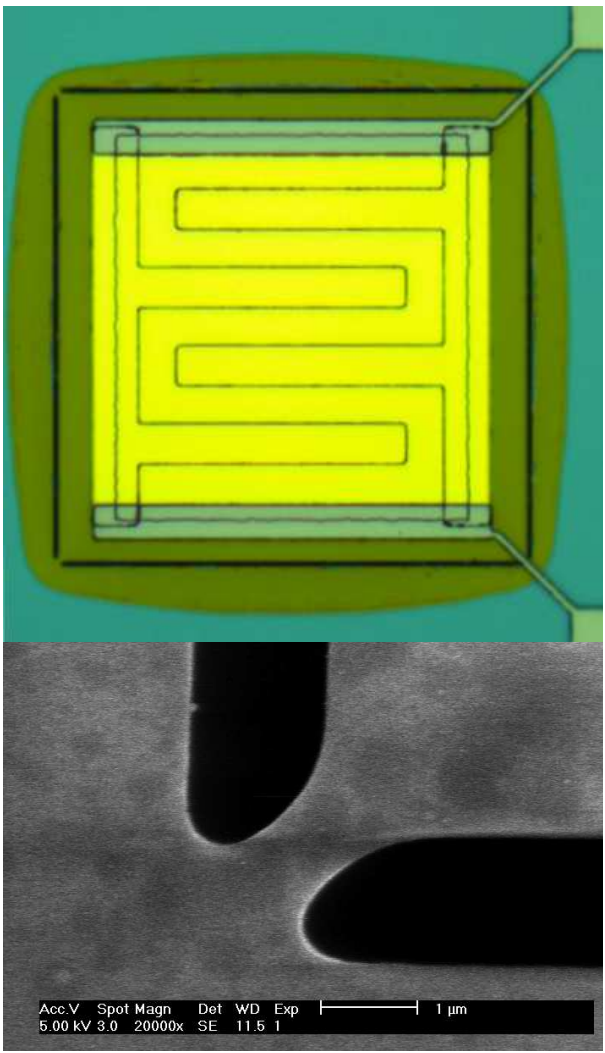}
\par\end{centering}
\caption{(Top) A photograph of one of the few-mode ballistic MoAu TESs; (bottom) an SEM image of one of the dielectric support legs, with a 1 $\mu$m rule for scale.
\label{fig6}}
\end{figure}

TESs were manufactured with exceedingly small amorphous SiN$_{\rm x}$ support legs: 0.7 to 1.0~$\mu$m wide and 1.0 to 4.0~$\mu$m long. A typical TES is shown in Fig.\ref{fig6}. Scanning Electron Microscopy (SEM) was used to measure the lengths and widths of the legs for each device. We took the smallest width to be the actual width, although
the width varied slightly along the length. The devices were fabricated in the form of 16-element arrays, and we did not have any failures, either because of broken legs, or because the Nb tracks were open circuit.

\section{\label{sec:3}Results}

The negative electro-thermal feedback of a voltage-biased TES causes the temperature of the island to essentially stay constant, and therefore any change in the DC power dissipated in the bilayer must correspond to changes in the power flow down the legs. By measuring dissipated power as a function of bath temperature, we were able to determine $P(T_{h},T_{c})$ as a function of $T_{c}$.

Figure \ref{fig4} shows a set of measured power-flow curves. In each plot, the measurements are shown as blue squares, and the model of Eq.~(\ref{eq:11}) as red lines. Careful inspection reveals that the curves are slightly different by virtue of the different geometries used. Remarkably, in all cases, the model based on the bulk elastic constants agrees exceedingly well with the measured values. We emphasize that there are no free parameters in the model, nor have we scaled the data, and therefore the thermal behaviour is being calculated from first principles.

\begin{figure}[h]
\noindent \begin{centering}
\includegraphics[scale=1]{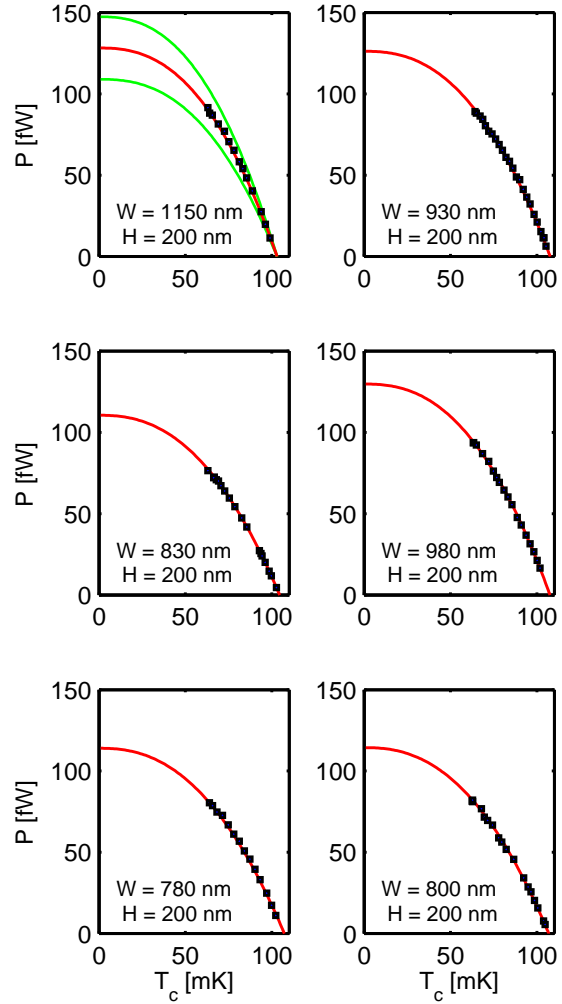}
\par\end{centering}
\caption{Net power flow, $P$, plotted against bath temperature $T_{c}$, for each of the TESs listed in Table \ref{table1}. The square points show measured data, and the red lines show the ballistic elastic phonon model.
The green lines in the first plot show $\pm$~15\% limits on the power flow.  
\label{fig7}}
\end{figure}

When designing TESs, it is usual to assume that the heat flow has the functional form \cite{Irwin2005}
\begin{equation}
P_{E}(T_{h},T_{c}) = K(T_{h}^{n}-T_{c}^{n})
\mathrm{,}
\label{eq:13}
\end{equation}
where $K$ and $n$ are constants. They depend on geometry, material, scattering lengths and temperature range, and so they have to be determined experimentally for each geometry considered, which is very time consuming. By fitting Eq.~(\ref{eq:13}) to the models in Fig. \ref{fig7} we were able to determine the values of $K$ and $n$ for the few-mode ballistic legs. Additionally, we were able to calculate the thermal conductances: $G={\partial P}/{\partial T_{h}}$. The results are listed in Table \ref{table1}. The length of each leg can only be regarded as being indicative, but crucially the thermal behaviour seems independent of length up to the largest value 3.9~$\mu$m. The fact that we do not see a $L^{-1}$ dependence indicates that the transport is ballistic. Future measurements will be undertaken to determine the actual value of the acoustic attenuation length, but that is not done here.

\begin{table*}
\noindent \begin{centering}
\begin{ruledtabular}
\begin{tabular}{ccccccc}

Width $\times$ Thickness (nm) & Length (nm) & $T_{h}$ (mK) & $n$ & $K$ (pW~K$^{-n}$) & $G$ (pW~K$^{-1}$)& $N$ ( $T_{c}=63$~mK)\tabularnewline
\hline
1150 x 200 & 1200 & 103.0 & 2.47 & 35.4 & 3.08 & 7.18\\
930 x 200 & 1100 & 107.6 & 2.43 & 28.3 & 2.85 & 6.37\\
830 x 200 & 1900 & 104.5 & 2.38 & 23.8 & 2.51 & 5.88\\
980 x 200 & 1700 & 107.5 & 2.44 & 30.2 & 2.95 & 6.58\\
780 x 200 & 2600 & 107.2 & 2.37 & 22.5 & 2.52 & 5.73\\
800 x 200 & 3900 & 106.8 & 2.37 & 23.2 & 2.54 & 5.80\\
\end{tabular}
\end{ruledtabular}
\par\end{centering}
\caption{Leg dimensions (width, thickness and length) for each of the 6 TESs tested, alongside corresponding
values of transition temperature $T_{h}$, empirical constants $n$ and $K$, thermal conductance $G$, and number of effective modes when $T_{c} = 63$ mK, $N$.}
\label{table1}
\end{table*}

The insert of Fig. \ref{fig8} shows the readout noise-power spectrum of one of the devices, and from this we calculated the NEP based on the power-to-current responsivity. The responsivity was determined through a
small signal model that was based on precise TES impedance measurements\cite{Figueroa-Feliciano2006,Lindeman2004,Goldie2009}.
The small-signal model was then used to calculate the contributions from the key noise sources as a function of readout frequency. These kinds of models have been reported previously and so we shall not describe them again here. The
key outcome is that the readout noise in Fig. \ref{fig8} is dominated by $1/f$ noise at $f<$10~Hz, phonon noise in the legs between $f=$ 10~Hz and $f=$ 50~Hz, `excess noise' between $f=$ 50~Hz and $f=$ 1~kHz, and Johnson noise at $f>1$~kHz. We will discuss the origin of the excess noise in Section~{\ref{sec:4}. The NEP of all of the ballistic devices was in the range 1.1 to 1.2~aW~Hz$^{-1/2}$ with a bath temperature of 63 mK. Figure \ref{fig8} shows as a red line, the NEP calculated on the basis of Eq.~(\ref{eq:13}). Thus, not only are the measured devices phonon shot-noise limited in the appropriate range, but the level of the phonon noise can be calculated on the basis of the elastic noise-wave model.

\begin{figure}[h]
\noindent \begin{centering}
\includegraphics{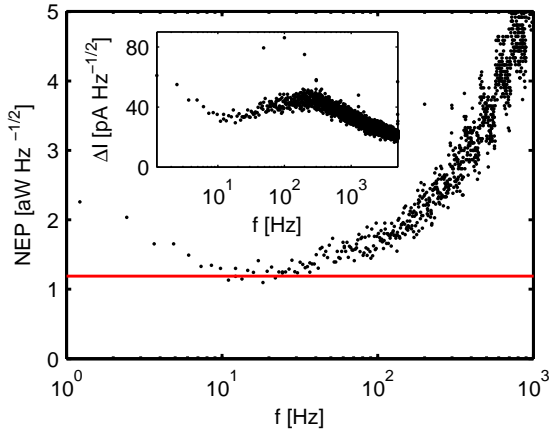}
\par\end{centering}
\caption{NEP against readout frequency, $f$, where the black points are measured data and the red line is calculated using the ballistic elastic phonon model. The insert shows the raw readout noise power spectrum. $T_{h} = 104.5$ mK$, T_{c} = 63.0$ mK, and the leg had cross sectional dimensions of 200~nm $\times$ 830~nm.
\label{fig8}}
\end{figure}

It is interesting to consider the number of ballistic modes that are responsible for carrying power. The presence of modes with non-zero cut-off frequencies introduces channels that can only transfer some fraction of the total emission spectrum of the reservoirs. It is convenient to define an effective number of modes according to
\begin{equation}
N(T_{h},\, T_{c}) = \frac{P(T_{h},T_{c})}{\int_{0}^{\infty} {h\nu} \left[{f(\nu,T_{h})}-{f(\nu,T_{c})}\right] d\nu}
\mathrm{.}
\label{eq:14}
\end{equation}
For temperatures $T_{h}$ and $T_{c}$ that are sufficiently low so that only the 4 lowest-order modes carry power $N(T_{h},\, T_{c}) = 4.0$, which is the lowest achievable limit. As the temperature is increased, additional modes make partial contributions, and $N(T_{h},\, T_{c})$ takes on non-integer values. Table~\ref{table1} lists $N(T_{h},\, T_{c})$, when $T_{c} = 63$~mK, for one leg of each of the ballistic TESs. In each case, effectively 6 to 7 elastic modes are responsible for carrying power, which is close to the limit of 4. Although all of our TESs operate close to the limit, the minimum value is not quite achieved: a leg having having a cross section of  200~nm $\times$ 400~nm would be needed to obtain the strict limit with $T_{c}=$ 63~mK. These observations are consistent with the findings of Schwab \cite{Schwab2000}.

The value of $n$ in Eq.~(\ref{eq:13}) is often regarded as indicating the dimensionality of the transport \cite{Hoevers2005,Karvonen2009,Leivo1998}, particularly in the ballistic regime. Roughly speaking, for $d$-dimensional transport, one might expect $n= d+1$. In the low-temperature one-dimensional limit care is required, however, because although the 4 lowest-order modes do indeed give $n=2$, an increasing number of modes cut on as the temperature is increased, giving rise to $n>2$. The total power transmitted can be written $P(T_{h},\, T_{c}) = N(T_{h},\, T_{c}) B(T_{h},\, T_{c})$, where $B(T_{h,\,}T_{c})$ is the power transmitted by one effective mode. If $N(T_{h},\, T_{c})$ is a constant, then $n=2$.  Generally $N(T_{h},\, T_{c})$ increases if either $T_{h}$ or $T_{c}$ increases, even though the power decreases as $T_{c} \rightarrow T_{h}$. Figure \ref{fig9} shows $N(T_{h},\, T_{c})$ as a function of $T_{c}$ for a SiN$_{\rm x}$  leg having a cross section of 1~$\mu$m $\times$ 1~$\mu$m, and the insert shows $B(T_{h},\, T_{c})$.
In this case we find, even for notionally one-dimensional transport, that for all four legs $n=3.08$, $K=$ 188.8~pW~K$^{-n}$, and $G=$ 5.3~pW~K$^{-1}$. The key point is that at low temperatures the value of $n$ is not a good indicator of dimensionality, and the situation becomes even more uncertain when scattering is included.

\begin{figure}[h]
\noindent \begin{centering}
\includegraphics[scale=1]{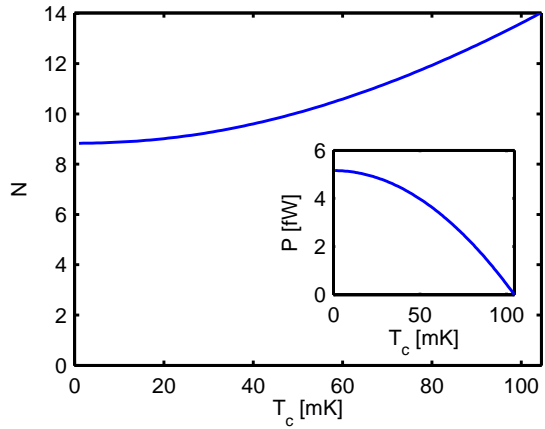}
\par\end{centering}
\caption{Effective number of modes, $N(T_{h},\, T_{c})$, plotted against the bath temperature, $T_{c}$, for a SiN$_{\rm x}$ leg having a cross section of 1~$\mu$m $\times$ 1~$\mu$m. $n=3.08$ in this case. The insert shows the power transmitted by a single mode as a function of bath temperature.
\label{fig9}}
\end{figure}

\section{\label{sec:4}Discussion}

We have fabricated a range of MoAu TESs whose SiN$_{\rm x}$ support legs only transmit of order 6 ballistic elastic modes. We have made a large number of these devices with 100\% yield, and the legs are strong and stiff. The TESs exhibit near ideal behaviour, and we have been able to calculate the power flow along the legs, and therefore conductance, as a function of the termination temperatures solely from knowledge of the bulk elastic constants of the material. It has not been necessary to carry out measurements to determine the the values of $K$ and $n$ in a parametric fit. The measured NEP is in excellent agreement with the ballistic thermal fluctuation model, and indicates that the noise is dominated by phonon shot noise at $T_{h}=$~100~mK. The measured NEP was 1.1 to 1.2~aW~Hz$^{-1/2}$, which is already suitable for ground-based astronomy, and would be ideal for space-borne submillimetre-wave CMB telescopes such as PRISM \cite{PRISM}.

The small-signal conductances of the few-mode ballistic TESs seem to be fully accounted for by the different leg geometries used.  All of the plots shown in Fig.~\ref{fig7} correspond to legs having different cross sections, and so it is difficult to quantify the variation in $G$ that would be expected for notionally identical devices. In the first plot in Fig.~\ref{fig7} we show, as green lines, $\pm$~15\% limits on the power flow, which would be typical of conventional long legs where internal scattering is used to achieve conductances that scale as $L^{-1}$. Looking at all of the results in Fig.~\ref{fig7}, it seems that the random variation is at a much lower level than that usually seen, which indicates that array uniformity should be improved considerably. In the case of conventional devices, AFM and SEM
show that the level of variation is much greater than can be accounted for purely on the basis of device processing or surface roughness, and is probably caused by localisation, where phonons are trapped in the disorder of the material. This process is intrinsic to the material and cannot be overcome if one relies on internal scattering to determine conductance. By contrast, the few-mode ballistic TESs described here appear to overcome this difficulty, and show high levels of uniformity. We take this to indicate that the 4~$\mu$m long legs are sufficiently short that localized resonant scattering has been eliminated.

The most demanding applications, such as the cooled-aperture far-infrared space telescope SPICA, require TESs having an NEP of 0.2~aW~Hz$^{-1/2}$, but this is 25\% of the minimum NEP that can be attained with 4 legs each having 4 ballistic modes. The conductances of the few-mode legs tested show no length dependence, indicating that the acoustic attenuation length is much greater than 4~$\mu$m. Also, the dominant phonon wavelengths lie in the range 400~nm to 2~$\mu$m indicating that it should be possible to fabricate simple patterned phononic support structures that reduce the heat flow. Single-stage filters should be able to achieve this level of attenuation, and double-stage filters certainly can. The temptation is to create Fabry-Perot filters through the use of step discontinuities, either in width or height, but because the discontinuities would be relatively smooth on the scale size of a wavelength, we prefer to use a scheme based on tiny, micromachined two-arm interferometers. The interferometric filters could be used individually; placed in close proximity to act as a coherent two-stage filter; or placed more than an acoustic attenuation length apart in order to use diffusive scattering to achieve phase isolation. Because the scattering process that limits the thermal flux is truly elastic, we would expect the thermal fluctuation noise to fall accordingly.

The insert in Figure~\ref{fig8}} shows a hump in the readout noise spectrum at around $f=$ 200~Hz, and features of this kind are well known in the TES community \cite{Zhao2011,Beyer2011a}. The hump usually appears at the frequency where the electrothermal feedback roles off, and is associated with the phase change in loop gain at the corner point. These features can be modelled by adding an additional heat capacity coupled to the island through a low thermal conductance. It is essential to include the thermal fluctuation noise of the weak link in order to produce the hump in the readout noise spectrum. In the case of the ballistic devices described here, an additional heat capacity of 10~fJ~K$^{-1}$ is required that couples to the TES island through a conductance of 12~pW~K$^{-1}$. For comparison, the total heat capacity of the island is 24~fJ~K$^{-1}$. The additional heat capacity cannot be accounted for on the basis of theoretical calculations of what one would expect for the constituent films. In the case of long, narrow ($>$ 500~$\mu$m $\times$ $<$ 2.0~$\mu$m) legs, it has been suggested that the distributed heat capacity of the legs might account for this behaviour\cite{Khosropanah2012}, but in the case of our ballistic devices the heat capacity of the legs is negligible, and yet we still see a hump. TLSs account for the majority of the heat capacity in amorphous SiN$_{\rm x}$, and they are known to have complex dynamical behaviour at low frequencies. It is possible that the time-dependent heat capacity of SiN$_{\rm x}$, or surface states, may be responsible for the hump. Because, however, Si has very similar dispersion characteristics to SiN$_{\rm x}$, it should be possible to make phononic TES support structures using SoI (Silicon on Insulator) \cite{Silverberg2008,Rostem2012}, and this would reduce the heat capacity of the devices considerably, and possibly eliminate or at least reduce the hump seen in readout noise spectra.

Overall, we envisage a new generation of TESs based on few-mode transport through micromachined phononic support structures. The legs would be tiny, less than 1~$\mu$m wide and 5~$\mu$m long. If fabricated using SoI techniques, they would eliminate many of the problems associated with amorphous SiN$_{\rm x}$, and would have near ideal performance: ultra-low-noise, high optical packing, high levels of array uniformity, flat readout noise spectra, and fast time constants. They could be fabricated with either free-space meshed absorbers or with superconducting microstrip coupling.

\section*{Acknowledgements}

We would like to thank Michael Crane for assistance in cleanroom processing, Dennis Molloy for mechanical engineering,
and David Sawford for electronic and software engineering.  Djelal Osman acknowledges the receipt of an STFC PhD studentship.

\bibliography{bibliog}

\end{document}